\documentclass[aps,prb,reprint,groupedaddress]{revtex4-1}
\usepackage{graphicx}

\draft 

\def\sb{SmB$_{6}$}

\begin{document}


\title{Surface properties of \sb\ from x-ray photoelectron spectroscopy} 

\author{N. Heming$^{1}$, U. Treske$^{1}$, M. Knupfer$^{1}$,  B. B\"uchner$^{1,2}$, D. S. Inosov$^{2}$, N. Y. Shitsevalova$^{3}$, V. B. Filipov$^{3}$, S. Krause$^{4}$, and A. Koitzsch$^{1}$}

\affiliation
{$^{1}$Institute for Solid State Research, IFW-Dresden, P.O.Box 270116, D-01171Dresden, Germany\\
$^{2}$Institut f\"ur Festk\"orperphysik, Technische Universit\"at Dresden, 01062 Dresden, Germany\\
$^{3}$Institute for Problems of Material Sciences, Kiev, Ukraine\\
$^{4}$BESSY II, Albert-Einstein-Str. 15, 12489 Berlin, Germany\\}

\date{\today}

\begin{abstract}
We have investigated the properties of cleaved \sb\ single crystals by x-ray photoelectron spectroscopy.
At low temperatures and freshly cleaved samples a surface core level shift is observed which vanishes when the temperature is increased. A Sm valence between 2.5 - 2.6 is derived from the relative intensities of the Sm$^{2+}$ and Sm$^{3+}$ multiplets. The B/Sm intensity ratio obtained from the core levels is always larger than the stoichiometric value. Possible reasons for this deviation are discussed. The B 1s signal shows an unexpected complexity: an anomalous low energy component appears with increasing temperature and is assigned to the formation of a suboxide at the surface. While several interesting intrinsic and extrinsic properties of the \sb\ surface are elucidated in this manuscript no clear indication of a trivial mechanism for the prominent surface conductivity is found.

\end{abstract}


\maketitle 

\section{Introduction\\}
\sb\ is an exceptionally well known and yet highly controversial compound. It was the first example of a mixed valent material  and as such received a lot of attention. \cite{Vainshtein1965, Nickerson1971, Wachter1994} Mixed valency refers to an instability of the f-electron count. The f$^5$ and f$^6$ configurations are found simultaneously when a fast probe like photoemission is employed and are thought to fluctuate in time. This is only possible if the f-bands are cut by the Fermi energy. However, the system does not behave like a metal but shows signs of a gap opening at low temperatures. \cite{Allen1979, Travaglini1984} This gap opening has been explained by a Kondo-lattice type hybridization, where the flat f-bands mix with the conduction bands such that a hybridization gap opens up. Since Sm has an even count of f- and d-electrons, the Fermi energy happens to be situated within the gap hence making \sb\ the showcase of a Kondo (or better: mixed valent) insulator. Based on this assignment the observation of a saturation of the resistivity at temperatures well below the hybridization gap opening is unexpected and opens the stage for the current discussion upon the nature of this conducting in-gap channel.\cite{Allen1979, Cooley1995} For more than three decades this state was discussed e.g. as impurity band or surface contribution.\cite{Flachbart2001, Kebede1996} But no consensus could be reached and its origin remained essentially elusive. A theoretical proposal renewed the interest in \sb\ far beyond the f-electron community.\cite{Dzero2010} The in-gap state was proposed to be a topologically protected surface state, and hence \sb\ is possibly the first representation of a topological Kondo insulator.    

Since then elaborate transport experiments clearly showed that the conductivity is due to a robust surface state with high charge density and low resistivity, difficult to reconcile with classic transport regimes. \cite{Wolgast2013, Kim2013} 

Several angle-resolved photoelectron spectroscopy (ARPES) studies have been made available recently focusing on the in-gap electronic structure. \cite{Miyazaki2012, Jiang2013, Xu2013, Neupane2013, Frantzekakis2013, Denlinger2013b, Zhu2013, Min2014} In-gap states cutting the Fermi energy are clearly identified. The majority assigns these states to the proposed topological nontrivial ones  based on circumstantial evidence. Also signatures of the expected spin texture have been found. \cite{Xu2014} However, distinctly different interpretations have been given \cite{Frantzekakis2013, Zhu2013} suggesting the possibility of surface off-stoichiometry (as a means to change the surface Fermi energy) and the existence of dangling boron surface bonds giving rise to a surface band. 

Another natural experimental probe for such surface phenomena is scanning tunneling microscopy (STM). \sb\ does not possess a natural cleavage plane. The surface of cleaved single crystals is therefore prone to reconstruction and disorder. The surface consists of different types of terminations on microscopic length scales. Nevertheless signatures of the hybridization gap and residual conductivity indicative of surface conduction have been found. \cite{Yee2013, Roessler2014, Ruan2014}

It is fair to say that to date the topological nature of the surface state is not unanimously confirmed and a more conventional surface phenomenon that leads to a conducting layer cannot be ultimately excluded. In this situation a notorious complication of f-electron systems demands extraordinary attention:  it is well known that f-systems tend to surface valence instabilities, both as an intrinsic process due to the breaking of translational symmetry and the reduced coordination number of f-ions at the surface but also due to rapid surface degradation e.g. by oxidation and consecutive change of the surface f-count.\cite{Treske2014_2, Wertheim1978}
Moreover the surface along the (001) direction is polar, the crystal structure consists of positively charged Sm layers alternating with negatively charged B-layers. We face a situation where the subject of our interest is a surface phenomenon which occurs in a material prone to surface changes and the fingerprints of its quantum nature are only accessible via surface sensitive methods like ARPES. Hence it is mandatory to study the surface in as much detail as possible. We do this here by applying soft x-ray photoelectron spectroscopy. We focus on the Sm valence, the chemical state of boron and the surface stoichiometry and search for anomalous behavior. 

\section{Experimental\\}
Soft x-ray photoelectron spectroscopy measurements were performed at UE52-PGM beamline at Berlin Synchrotron Facility (BESSY II) in the excitation energy range of 330\,eV to 1400\,eV using linearly polarized light. Since the inelastic mean free path of the photon excited electrons depends on their kinetic energy, the surface sensitivity of the photoemission spectra is increased by reducing the excitation energy. Spectra were recorded by a Scienta R4000 photoelectron analyzer. The total energy resolution in the more surface sensitive case ($h\nu$ = 330\,eV) was about $\Delta$$E$ = 100\,meV and in the more bulk sensitive case ($h\nu$ = 1400\,eV) $\Delta$$E$ = 400\,meV. The beam spot size is  about 0.25 $\times$ 0.6 mm$^2$, the sample surface 1 $\times$ 1 mm$^2$.

Single crystal samples of \sb\ as well as CeB$_6$ and LaB$_6$ were prepared by floating zone method as described elsewhere.\cite{Friemel2012} Before the measurement, the crystals were oriented by Laue diffraction and on the outside of the sample small notches were cut in order to create a predetermined cleavage plane along the (001) direction. After that the samples were cooled in ultrahigh vacuum with a base pressure of 1 $\times$ 10$^{-10}$ mbar by a He flow cryostat to 17 K and cleaved for the measurement.

\section{Results and Discussion\\}
\subsection{Valence band\\}
The valence of Sm in \sb\ has been evaluated by XPS already in the 70's. \cite{Chazalviel1976, Allen1980} Fig. 1 shows the valence band measured at freshly cleaved sample at T = 17 K with h$\nu$ = 330\,eV. The spectrum is dominated by intense Sm 4f related structures. The Sm 5d and B 2p states are suppressed due to  low photoemission cross sections and their delocalized nature. The Sm$^{2+}$ (f$^{6}$) and Sm$^{3+}$ (f$^{5}$) configurations are well separated from each other and form extended multiplet structures. It is the observation of these two configurations in the same spectra which characterizes \sb\ as being mixed valent. The shape of the multiplets can be derived from tabulated values for their relative weight and energy separation. \cite{Gerken1983}

\begin{figure}[h!]
\includegraphics[width=0.9\linewidth]{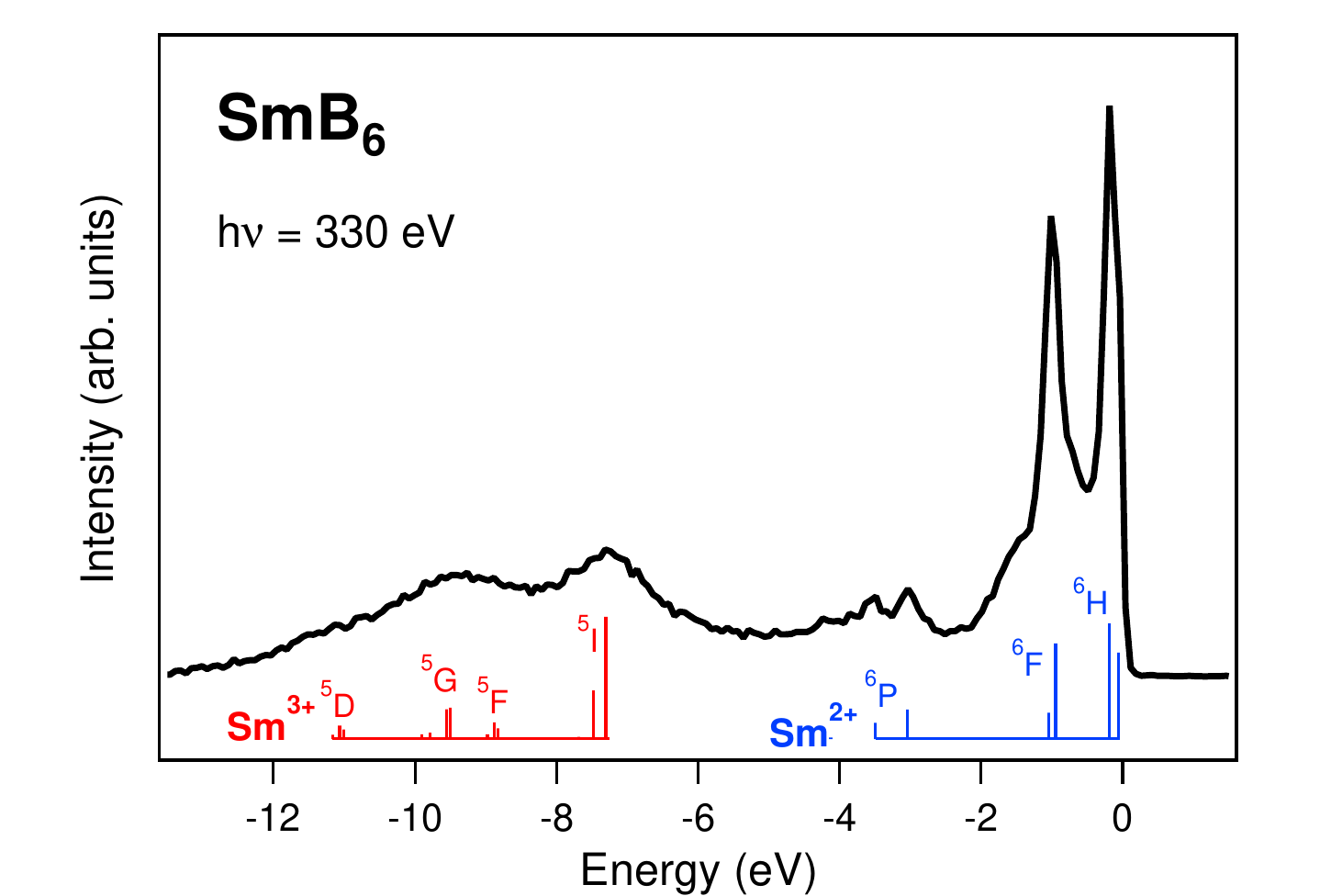}%
\caption{Valence band of freshly cleaved \sb\ at T = 17\,K taken with a photon energy of h$\nu$ = 330\,eV compared to multiplet states derived from tabulated values.\cite{Gerken1983} \label{}}%
\end{figure}

Generally the correspondence between multiplet theory and experiment is satisfactory except for a high energy shoulder at $E$ $\approx$ 1.5\,eV. In fact, this shoulder is part of a broader structure centered at $E$ $\approx$ 1\,eV. It has been reported before by Zhu et al. \cite{Zhu2013} and attributed to a B 2p related surface state. The intensity of the latter decreases with time due to suggested surface self annealing. Denlinger $et\ al.$, on the other hand, attributed it to a Sm$^{2+}$ related surface shift.\cite{Denlinger2013} We have observed that this 1\,eV feature depends strongly on the cleavage. Fig. 2 presents a comparison of the valence band shown in Fig. 1 (sample 1)  with another cleavage under very similar conditions (sample 2). The shoulder structure is absent but the Sm$^{3+}$ related multiplets match very well. Subtracting both spectra from each other results in the blue difference spectrum, which itself clearly resembles the Sm$^{2+}$ part of the valence band. 

\begin{figure}[h!]
\includegraphics[width=0.9\linewidth]{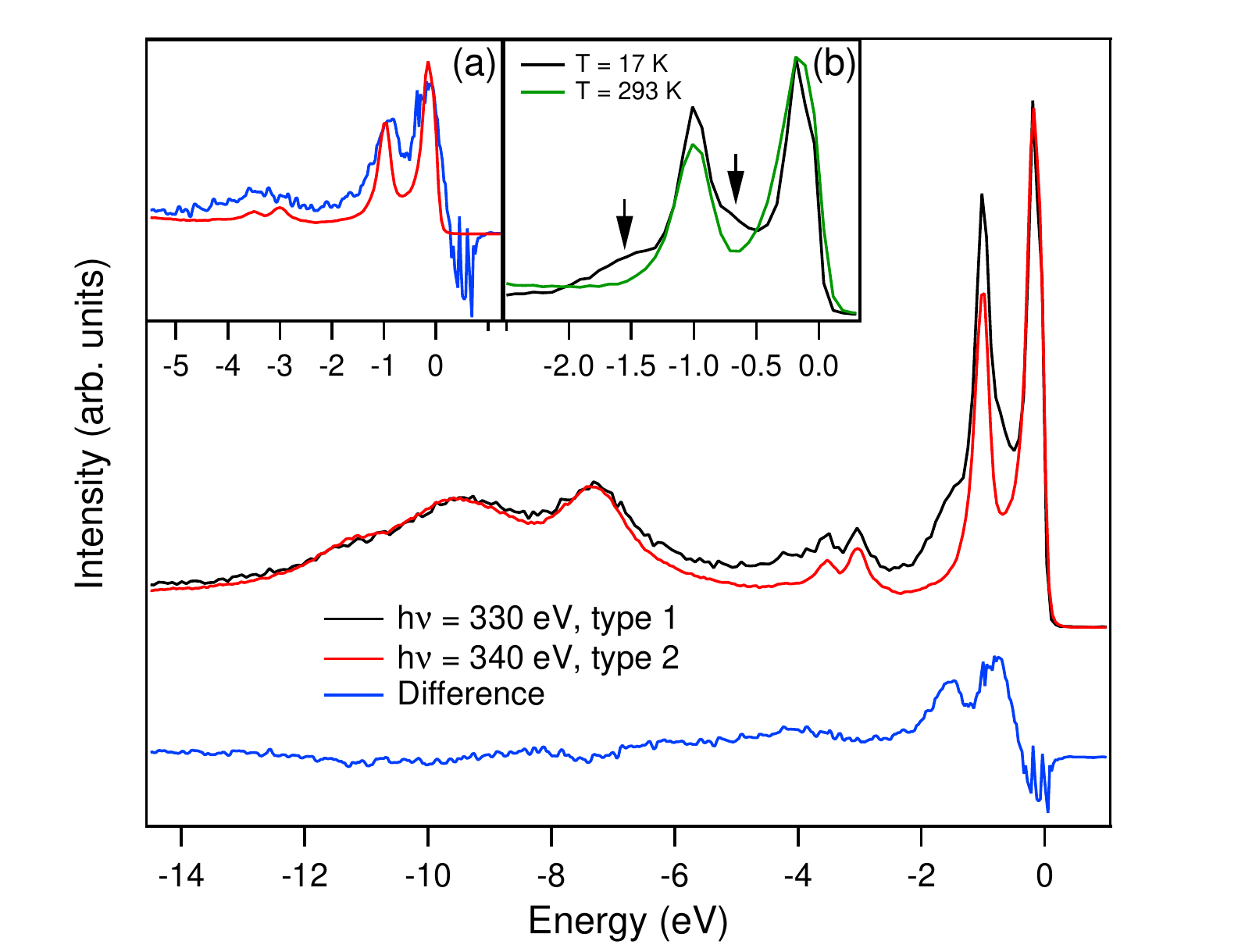}%
\caption{Comparison of sample 1 and sample 2 valence bands and their difference. Inset (a): Comparison of sample 2 valence band and difference spectrum shifted by $\Delta$$E$ = 0.6\,eV. Inset (b): Temperature dependence of the sample 1 valence band, showing the reduction of the surface state emission at room temperature.\label{}}%
\end{figure}

Inset (a) of Fig. 2 presents a normalized and energy shifted comparison of the difference spectrum with the sample 2 Sm$^{2+}$ signal. The agreement is convincing. Inset (b) compares the low temperature measurement of sample 1 right after the cleavage with a later room temperature measurement. Typical for surface states the feature vanishes with time and temperature. Therefore the 1\,eV feature can be identified as a surface related shift of the Sm$^{2+}$ multiplet by $\Delta$$E$ = 0.6\,eV. Such a shift is absent for the Sm$^{3+}$ multiplet. Also Sm$^{2+}$ has a larger relative spectral weight for sample 1 as compared to sample 2. This suggests that the sample 1 surface has an instability towards the 2+ state.

Fig.\,3 compares valence bands of sample 1 and sample 2 taken with different photon energy. The two photon energies entail different inelastic mean free paths (IMFP), i.e. different information depths with the 330\,eV measurement being  more surface sensitive. The inset of Fig. 3 shows a dependence of the IMFP on the kinetic electron energy calculated via the TPP-2M formalism. \cite{Tanuma2003}

\begin{figure}[h!]
\includegraphics[width=0.9\linewidth]{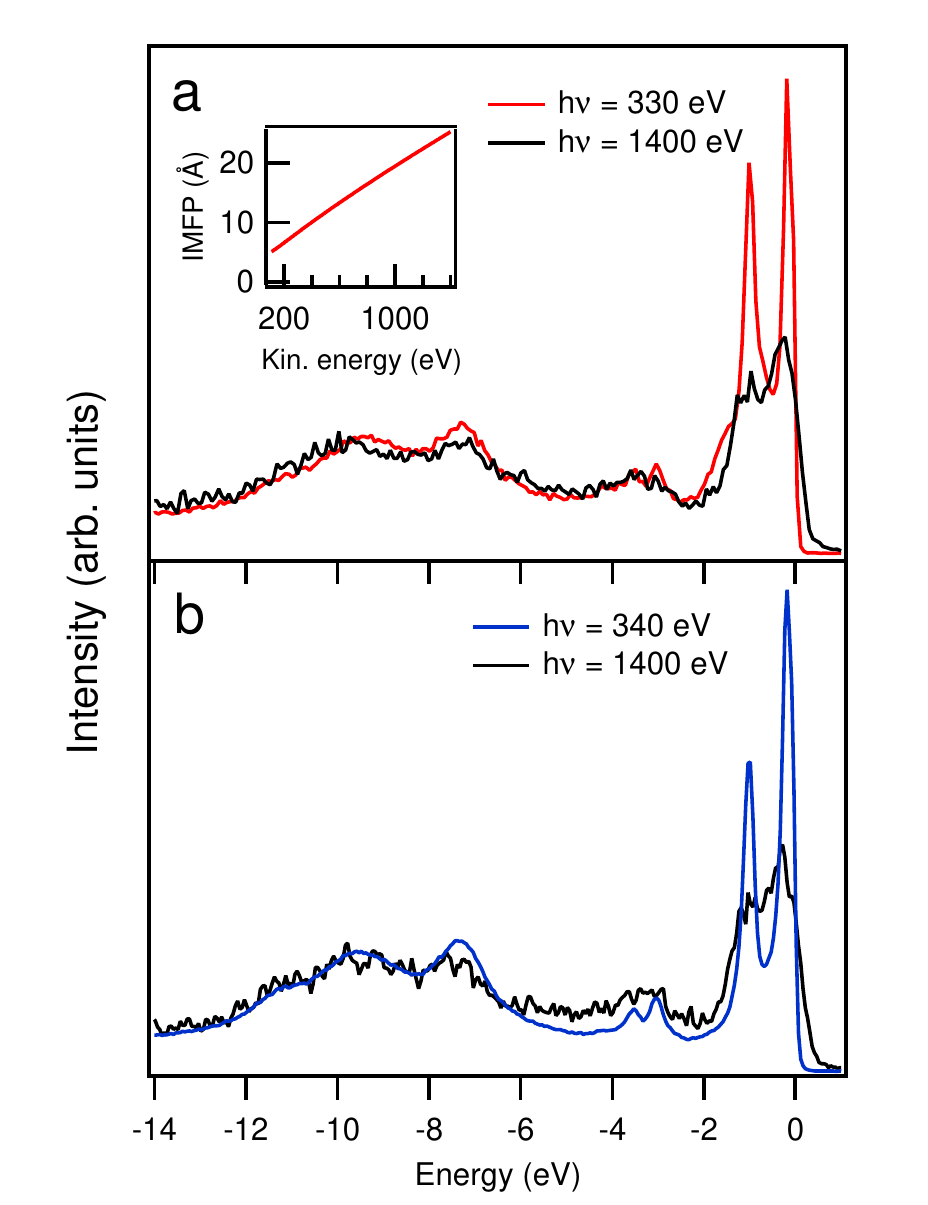}%
\caption{Comparison of (a) sample 1 and (b) sample 2 valence bands taken with different photon energies at low temperature to vary the depth sensitivity. Inset: Inelastic mean free path (IMFP) as a function of electron kinetic energy.\cite{Tanuma2003} \label{}}%
\end{figure}

From the intensity ratio of the Sm$^{2+}$ and Sm$^{3+}$ multiplets the effective Sm valence can be derived. Table 1 summarizes the results for sample 1 and 2 obtained by standard fitting procedures (not shown). The sample 1 surface shows a decreased surface valence in contrast to sample 2. Surface valence changes are well-known for rare earth materials and in particular for Sm metal and its compounds, e.g. elemental Sm, which shows  Sm$^{3+}$ in the bulk and Sm$^{2+}$ at the surface.\cite{Wertheim1978} Here a similar effect is observed. The values in table 1 are in good agreement with previous bulk measurements (e.g. 2.6 from Moessbauer spectroscopy\cite{Cohen1970}, 2.5 from bulk sensitive x-ray absorption spectroscopy \cite{Mizumaki2009}).    

 \begin{table}
 \caption{\label{tab:Table1} Sm valence extracted from the valence bands in Fig. 3.}
 \begin{ruledtabular}
 \begin{tabular}{ccc}
h$\nu$  & sample 1 & sample 2\\
\hline
330 / 340\,eV & 2.52 & 2.59 \\
1400\,eV & 2.59 & 2.55 \\
 \end{tabular}
 \end{ruledtabular}
 \end{table}

\subsection{Stoichiometry\\}
The surface termination of hexaborides has been systematically studied, motivated by the exceptionally low work function of LaB$_6$ and its use as a thermionic emitter.\cite{Trenary2012} LaB$_6$ has been found to be metal terminated,\cite{Aono1979} for \sb\ partial or full Sm termination has been suggested.\cite{Aono1979, Futamoto1980} However, these results concern non-cleaved surfaces, which were polished and underwent e.g. heating cycles before measurement, so that the surface reached equilibrium conditions. For cleaved single crystals  considered here, which are the basis for modern ARPES and STM measurements, a boron rich surface has been observed recently by XPS. \cite{Yee2013}

\begin{figure}[h!]
\includegraphics[width=1\linewidth]{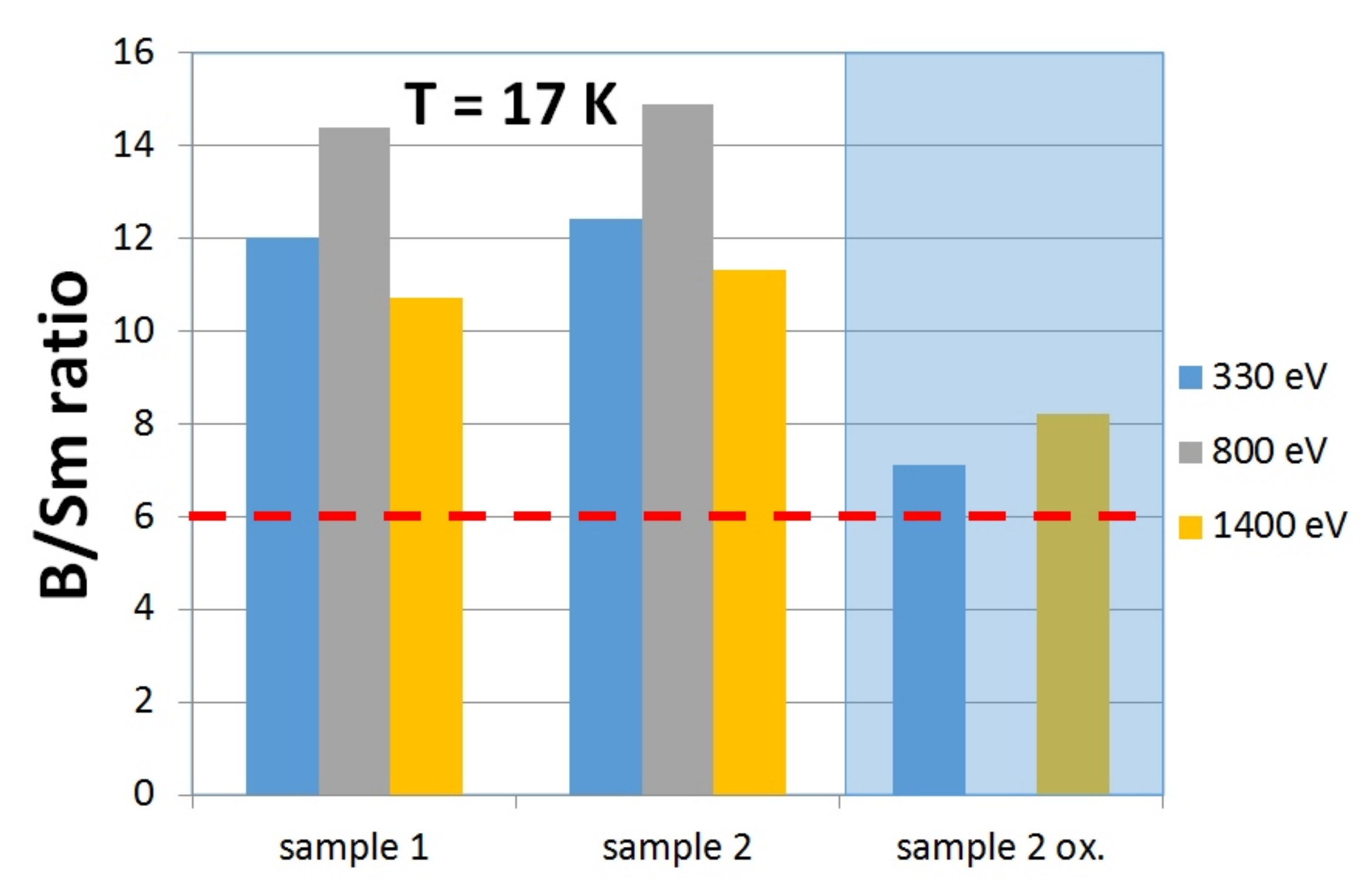}%
\caption{Stoichiometry extracted from the B 1s and Sm 4d core levels for different photon energies and samples. The blue shaded area on the right side shows the values for sample 2 after exposure to ambient atmosphere. The red dashed line represents the nominal stoichiometric value.\label{}}%
\end{figure}

Fig. 4 shows the B/Sm concentration ratio for two cleavages at low T and normal emission extracted from the B 1s and Sm 4d core levels after Shirley background subtraction and correction for the photoemission cross section and asymmetry parameter.\cite{Yeh1985} Differences of the transmission function and the IMFP of the B 1s and Sm 4d core levels have been neglected since the two lines are sufficiently close together in energy. Again, we applied different photon energies to vary the surface sensitivity. An estimated error of 15$\%$ of the final B/Sm values, mainly arising from uncertainties of the background subtraction, should be considered. 

Several interesting observations can be made from Fig. 4: i) the B/Sm ratio is significantly larger than nominal stoichiometric value of six, ii) it decreases drastically after exposure to ambient conditions, iii) no monotonous dependence of the B/Sm ratio on photon energy is found, the 800\,eV measurement delivers the largest value.

First we discuss the origin of the surprisingly large B/Sm value and whether it is indicative of a surface boron excess. Since photoemission spectroscopy is a very surface sensitive technique, the stoichiometry obtained might depend on the cleavage plane.
In fact a cleavage plane along (001), which delivers a boron termination, exists: namely a disruption of the B$_6$ octahedra. All other cleavage planes would result in Sm or B terminations mirroring on average the bulk stoichiometry. Indeed the surface consisting of disrupted B$_6$ octahedra has been observed by STM, although for crystals cleaved at room temperature.\cite{Ruan2014}
In order to estimate an upper limit of the stoichiometry that can be expected from such a boron termination we consider the following model: the bulk \sb\ is approximated by homogeneously mixed B and Sm with a concentration ratio of 6:1. Additionally, a boron top layer is assumed with d = 1.2 \AA, i.e. about half the thickness of a full boron layer. The XPS stoichiometry can then be approximated by B/Sm 	= 6/$e^{-d/\lambda}$. Even for $h\nu$ = 330\,eV with $\lambda$ $\approx$ 6 \AA\ this results in B/Sm = 7.3, which is still far below the experimental values. Also there is no indication of the presumed surface termination in the low temperature STM reports. An explanation relying purely on the type of termination is therefore ruled out.

The measurements in Fig.4 have been performed at normal emission. For single crystals an angle dependence can occur due to diffraction of the outgoing electron. Fig. 5a shows the B/Sm ratio as a function of the polar angle taken with a laboratory Al K$\alpha$ source (h$\nu = 1486.6$\,eV) at T = 20 K. Indeed a significant dependence is found. Diffraction maxima due to forward scattering are expected  at 0$^{\circ}$ and 45$^{\circ}$ in this cubic structure. Fig.5a is consistent with the assumption that B is a more efficient forward scatterer than Sm due to its octahedral configuration. It also indicates that the cleaved surface is of good crystalline quality.
Apart from the emission angle the diffraction depends also on the photon energy, which might explain the maximum of the B/Sm ratio for h$\nu$ = 800\,eV. The average of B/Sm from Fig. 5a is around 9, which is still larger then 6. For a more accurate value the range of the polar integration has to be enlarged and complemented by azimuthal scans.
 
Lastly, also surface reconstruction can influence the B/Sm ratio seen by photoemission. The presence of reconstructions of various types is well documented by STM. \cite{Yee2013, Roessler2014} R{\"o}{\ss}ler et al. report the appearance of chain like Sm structures on top of the boron planes. It seems possible that these Sm chains attenuate the boron signal less efficient than the original 1 $\times$ 1 Sm planes, and therefore effectively the B/Sm could increase. While the last point is a speculation it is safe to assert, that the standard IMFP formalism employed explicitly and implicitly above comes to its limits under these circumstances and should be substituted by more involved methods \cite{Werner2001}, which is beyond the scope of this study.  

We conclude from the above arguments that the large B/Sm ratio observed can be a consequence of diffraction processes and/or surface reconstruction and is not necessarily related to a surface boron overstoichiometry. However, the latter cannot be excluded and remains a natural possibility. Yee at al. do find the prevalence of a boron rich disordered surface where Sm terminations are underrepresented. \cite{Yee2013} Why exactly such a termination would be favorable and how the reconstruction looks microscopically remains to be clarified.  

\begin{figure}[h!]
\includegraphics[width=0.9\linewidth]{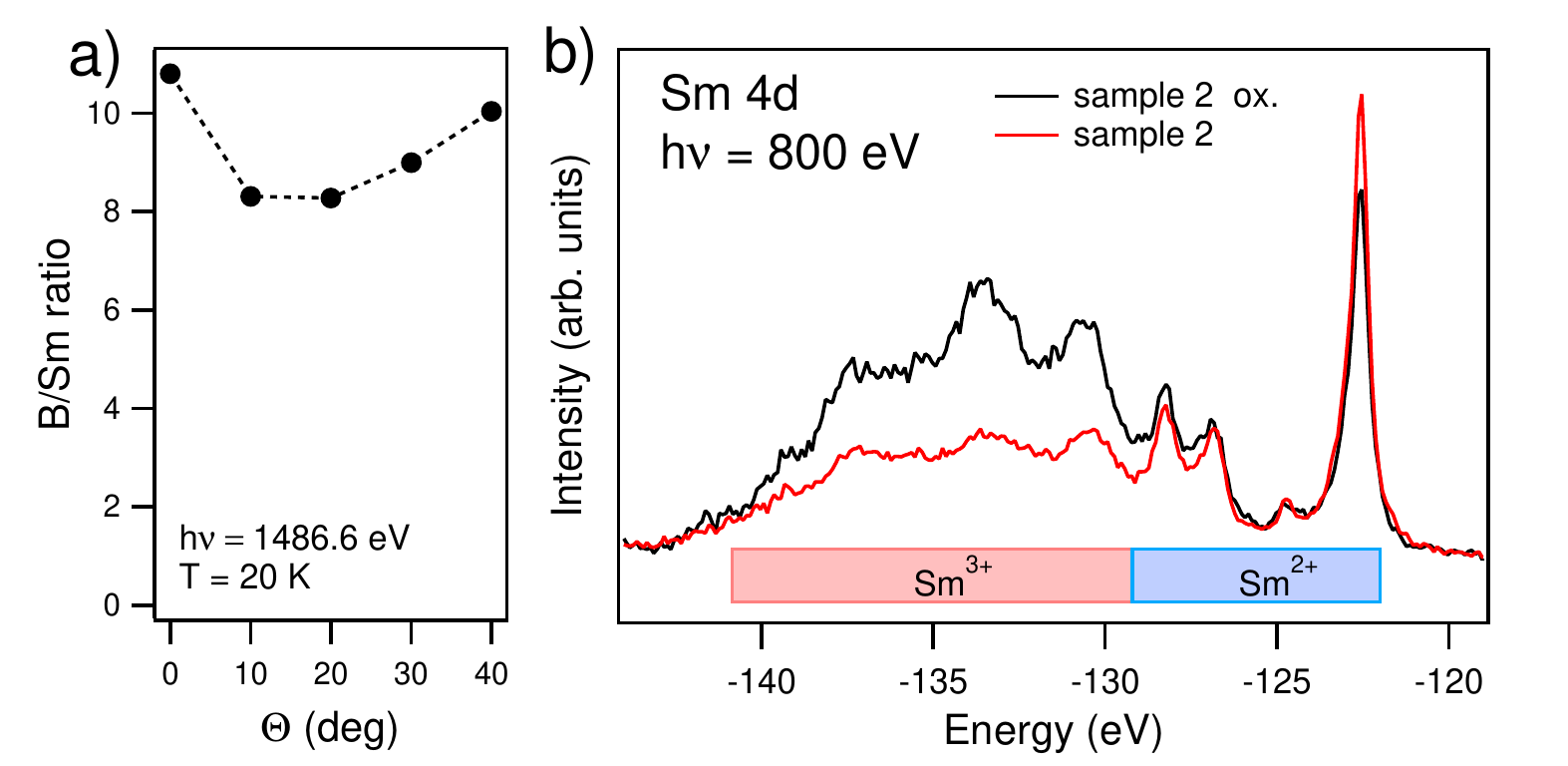}%
\caption{a) B/Sm intensity ratio extracted from XPS (see text) as a function of the polar angle. b) Comparison of the Sm 4d line right after cleavage and after exposure to ambient atmosphere. The lineshape is dominated by Sm$^{3+}$ and Sm$^{2+}$ related multiplet structures. \label{}}%
\end{figure}

Photoemission and STM studies are carried out under UHV conditions whereas the transport properties are usually investigated under ambient atmosphere. In order to crosscheck the implications of this difference we exposed one of the in-situ cleaved samples to air before transferring it back to the vacuum chamber (see Fig. 4). The stoichiometry differs significantly: the boron surplus is reduced. Fig. 5b presents a comparison of the Sm 4d lines normalized to the B 1s signal at higher energy. The Sm intensity increase of the air exposed sample is apparent. The line shape of the Sm 4d core level is similar to the Sm 4f emission in the valence band. It consists of Sm$^{2+}$ and Sm$^{3+}$ related multiplet structures at lower and higher energies respectively, which partly overlap. The intensity increase of the air exposed sample can be traced to an almost exclusive increase of the Sm$^{3+}$ region, which is most naturally associated with oxidation. The stoichiometry change with respect to the as cleaved surface may be then be associated with a Sm surface segregation driven by the surface oxidation potential or, alternatively, a change of the diffraction pattern.

\begin{figure}[h!]
\includegraphics[width=0.9\linewidth]{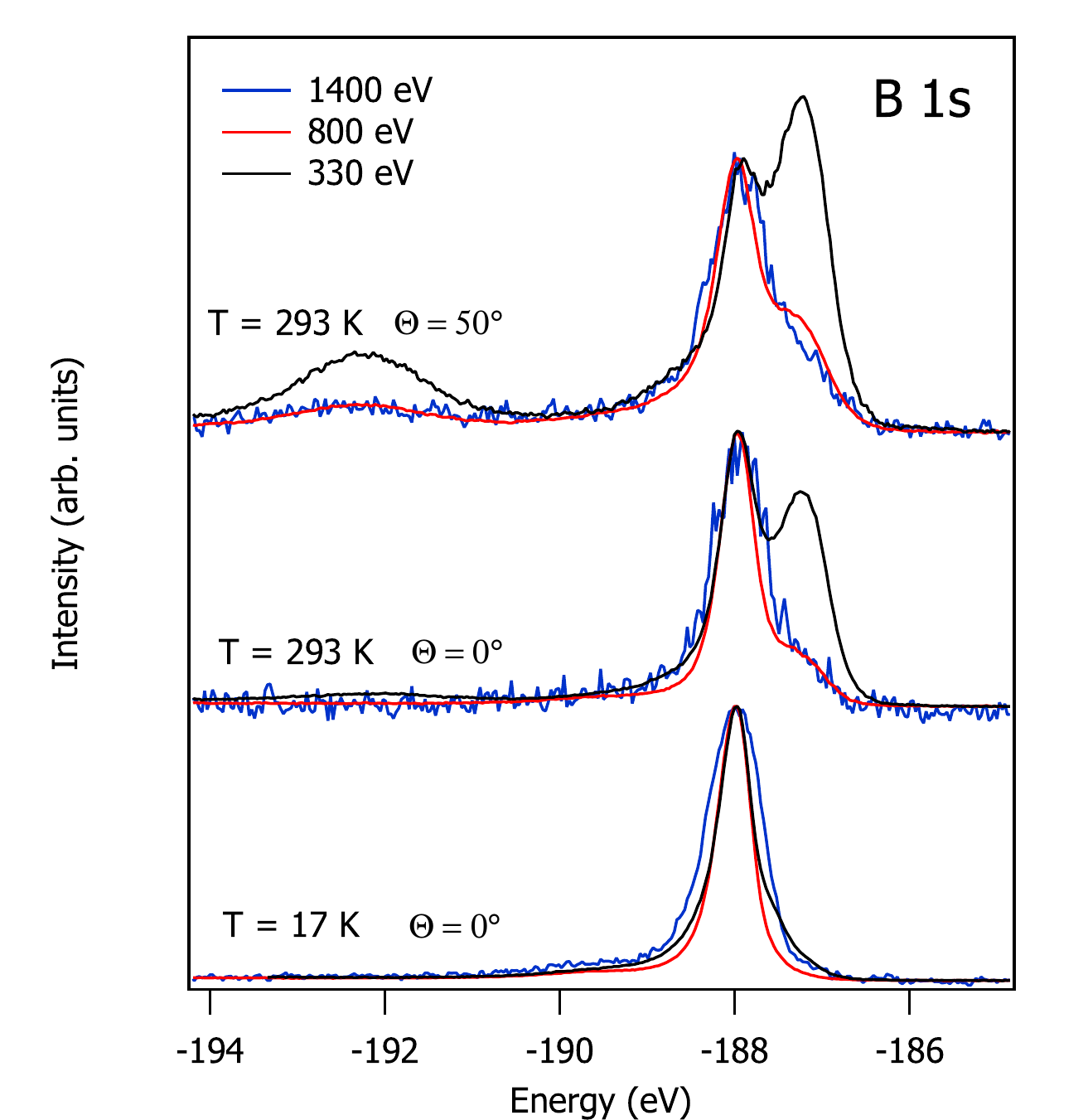}%
\caption{Boron 1s line as a function of photon energy, temperature and emission angle. \label{}}%
\end{figure}

\subsection{B 1s\\}
It is important to clarify also the chemical state of the surface boron. Fig. 6 shows the B 1s line measured with different photon energies and under different conditions. At low temperatures the B 1s line consists of a single sharp peak at $E$ = -187.9\,eV. For the most surface sensitive measurement ($h\nu$ = 330\,eV), a small low energy shoulder is observed. This shoulder increases strongly as the temperature is increased. The photon energy dependence clearly reveals its surface character. To enhance further the surface sensitivity, we varied the emission angle, which results in an increase of the low energy feature and the appearance of another peak at $E$ = -192.3\,eV. The latter can be identified as boron oxide (B$_2$O$_3$). The assignment of the low energy feature is less clear. A previous study on other hexaborides found a small low energy shoulder as well by fitting the B 1s line taken with Al K$\alpha$ excitation ($h\nu$ = 1486.6\,eV). It was attributed to a surface boron layer with reduced binding energy due to changing Madelung potentials at the surface, in particular the absence of half of the positively charged rare earth ions in the local environment of the B-octahedra. \cite{Patil2011} However this interpretation seems incompatible with the temperature dependence in Fig. 6. The effect is expected to be strongest right after cleavage and not after some time and temperature increase, when the surface had the possibility to relax. The same holds true for the related assumption of an energy shift due to the polar character of the surface. 

Fig. 7 presents the B 1s line for other hexaborides namely LaB$_6$ and CeB$_6$. Both show similar low energy components as \sb. This observation is in favor of a universal process affecting the boron subsystem.
In the following we will concentrate on a scenario where the anomalous low energy component is the signature of a surface contamination due to residual background gases. The presence of such gases is supported by the growth of the B$_2$O$_3$ peak, which is surely a contamination. After the cleavage many boron bonds must be broken. The crystal structure possesses different boron bonds: inter-octahedral and intra-octahedral bonds. It is interesting that the suboxide B$_6$O has been reported with a low binding energy peak at E = -187.1\,eV, matching the low energy component. \cite{Moddeman1989, Ong2004} We assign therefore the B$_2$O$_3$ peak at $E$ = -192.3\,eV to the oxidation of dangling intra-octahedral bonds, i.e. disrupted B$_6$ octahedra, and the low energy feature to the formation of suboxides, saturating broken inter-octahedral bonds. The intensity ratio of the low and high energy oxidation peaks in Figs. 6 and 7 shows that most of the B$_6$ octahedra remains intact, which is not surprising because of their rigid, covalently bond nature.

\begin{figure}[h!]
\includegraphics[width=0.9\linewidth]{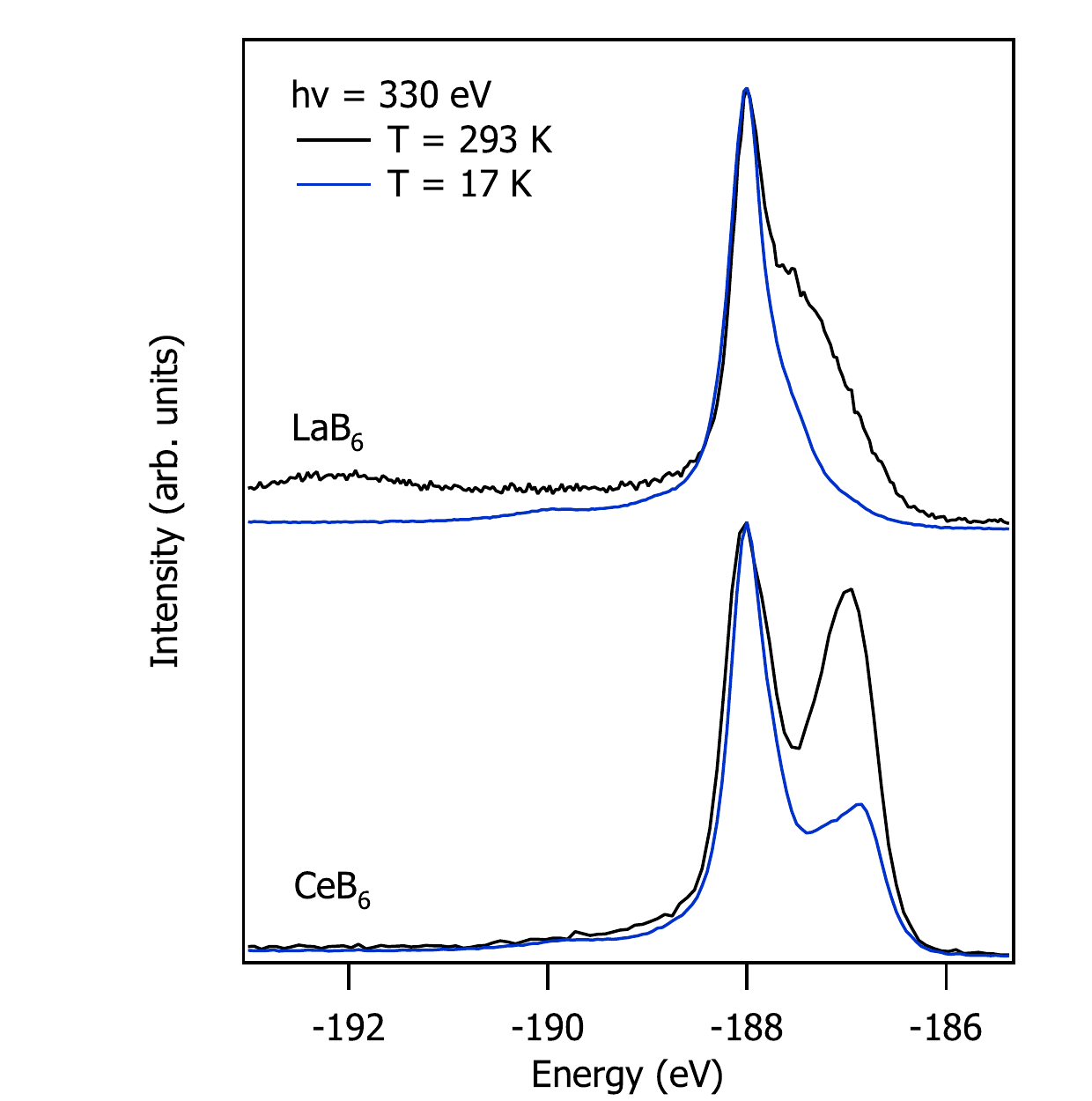}%
\caption{Boron 1s line for different hexaborides at low and room temperature. \label{}}%
\end{figure}

The room temperature spectra in Fig. 6 can be fitted by standard procedures to obtain the relative intensities of the boron components (not shown). With this information and the known IMFP (inset Fig. 3) the thickness of the boron surface layer responsible for the low energy shoulder and the oxide component can be evaluated by an exponential damping model:

\begin{eqnarray}
\frac{I_{\text{oxide}} + I_{\text{Low E}}}{I_{\text{Hexaboride}}}=\frac {1-e^{-d/\lambda}}{e^{-d/\lambda}} \label{eq:one}.
\end{eqnarray}

where $I_{\text{oxide}}$ refers to the intensity of the peak around $E = -192.3$\,eV, $I_{\text{Low E}}$ to the low energy shoulder at $E = -187.2$\,eV and $I_{\text{Hexaboride}}$ to the generic component at $E = -187.9$\,eV. $\lambda$ is the IMFP and $d$ the thickness of the surface layer. Eq. (1) implicitly assumes an equal number of boron atoms per unit volume for all the components. Fig. 8 shows the values of $\frac{I_{\text{oxide}} + I_{\text{Low E}}}{I_{\text{Hexaboride}}}$ extracted from Fig. 6 vs the IMFP calculated for the various photon energies and emission angles. The parameter $d$ is obtained from the best fit of the data to eq. (1). This results in $d = 2.5$ \AA, i.e. the first boron layer only. This is again consistent with the oxidation scenario discussed above, since this should only happen at the surface layer. 

\begin{figure}[h!]
\includegraphics[width=0.9\linewidth]{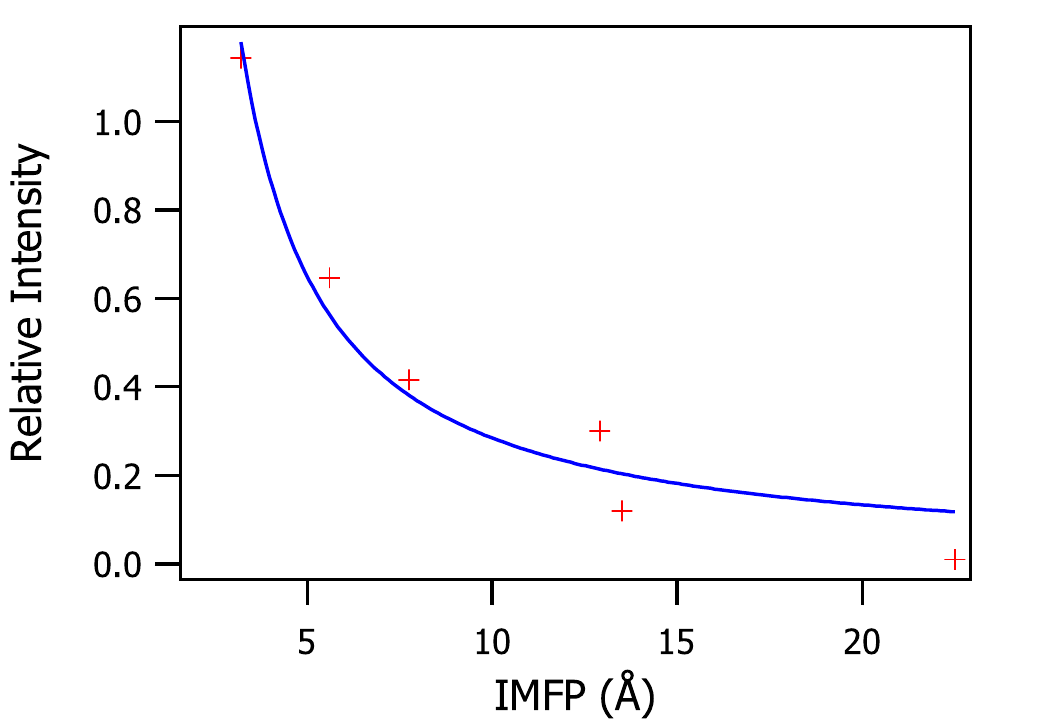}%
\caption{Markers: Intensity ratio of the boron surface components vs the generic hexaboride component as a function of the IMFP calculated from the various photon energies and emission angles used in Fig. 6. Solid line: Fit of the data with Eq. (1), see text. \label{}}%
\end{figure}

\section{Conclusions\\}
We observe a number of intrinsic and extrinsic phenomena at cleaved \sb\ surfaces: a surface core level shift for the  pristine surface at low temperatures associated with a Sm$^{2+}$ rich surface, which vanishes when the temperature is increased. A samarium valence of 2.5 - 2.6 is obtained from quantitative analysis of the valence band multiplets depending on the surface sensitivity of the measurement. Large B/Sm ratios are derived from quantitative analysis of the XPS data. They might be a consequence of efficient boron forward scattering and the reconstruction of the polar surface. More detailed XPD studies and model calculations for the photoelectron attenuation are necessary to clarify this point.

The B 1s line develops an unusual low energy component, which increases with increasing temperature. It is assigned to the formation of a boron suboxide at the surface layer, saturating broken inter-octahedral boron-boron bonds, apparently a universal process for all hexaborides. 

These findings do not offer an alternative, topological trivial scenario for the surface conductivity. However, it becomes clear that the surface of \sb\ features an unexpected complexity which warrants further investigations and has to be taken into account for the interpretation of surface sensitive measurements like ARPES and STM. 

\section{Acknowledgement\\}
D.S.I. acknowledges support from the German Research Foundation (DFG) under grant IN 209/3-1.

\bibliography{Ref_SmB6}

\end{document}